\documentclass[letter,traditabstract]{aa}

\usepackage{graphicx}

\usepackage{txfonts}

\begin{document} 

\title{Chemical properties in the most distant radio galaxy}

\author{K.~Matsuoka\inst{1,2} \and T.~Nagao\inst{2,3} \and R.~Maiolino\inst{4} \and A.~Marconi\inst{5} \and Y.~Taniguchi\inst{6}}

\institute{Graduate School of Science and Engineering, Ehime University, 2-5 Bunkyo-cho, Matsuyama 790-8577, Japan\\ \email{kenta@cosmos.phys.sci.ehime-u.ac.jp} \and Department of Astronomy, Kyoto University, Kitashirakawa-Oiwake-cho, Sakyo-ku, Kyoto 606-8502, Japan \and The Hakubi Project, Kyoto University, Yoshida-Ushinomiya-cho, Sakyo-ku, Kyoto 606-8302, Japan \and INAF -- Osservatorio Astrofisico di Roma, Via di Frascati 33, 00040 Monte Porzio Catone, Italy \and Dipartimento di Fisica e Astronomia, Universit\'a degli Studi di Firenze, Largo E. Fermi 2, 50125 Firenze, Italy \and Research Center for Space and Cosmic Evolution, Ehime University, 2-5 Bunkyo-cho, Matsuyama 790-8577, Japan}
 
\abstract{We present a deep optical spectrum of TN J0924$-$2201, the most distant radio galaxy at $z = 5.19$, obtained with FOCAS on the Subaru Telescope. We successfully detect, for the first time, the \ion{C}{iv}$\lambda$1549 emission line from the narrow-line region (NLR). In addition to the emission-line fluxes of Ly$\alpha$ and \ion{C}{iv}, we set upper limits on the \ion{N}{v} and \ion{He}{ii} emissions. We use these line detections and upper limits to constrain the chemical properties of TN J0924$-$2201. By comparing the observed emission-line flux ratios with photoionization models, we infer that the carbon-to-oxygen relative abundance is already [C/O] $> -0.5$ at a cosmic age of $\sim 1.1$ Gyr. This lower limit on [C/O] is higher than the ratio expected at the earliest phases of the galaxy chemical evolution, indicating that TN J0924$-$2201 has already experienced significant chemical evolution at $z = 5.19$.}


\maketitle


\section{Introduction} 

The chemical evolution of galaxies is closely related to their past star-formation history. Therefore the investigation of the chemical evolution of galaxies is crucial for understanding the formation history of galaxies. Observationally, the chemical evolution of galaxies has been studied by measuring the metallicity of galaxies at various redshifts. The gas-phase metallicity of star-forming galaxies in the local universe can be measured fairly easily by means of optical emission-line spectra (e.g., [\ion{O}{ii}]$\lambda$3727, H$\beta$, [\ion{O}{iii}]$\lambda$5007, H$\alpha$, and [\ion{N}{ii}]$\lambda$6584; see Nagao et al. 2006c and references therein). However, in the early universe at $z > 1$, these diagnostics are more difficult to use because these emission lines are often very faint at $z > 1$ and are redshifted into the near-infrared, where sensitive spectroscopic observations are more challenging.

Active galactic nuclei (AGNs), thanks to their high luminosity, offer an interesting alternative to investigate the metallicity in high-$z$ galaxies. The AGN are characterized by a wealth of bright emission lines in the wavelength range from rest-frame ultraviolet to the infrared, arising from gas clouds photoionized by their central engines. In particular, the ultraviolet lines, e.g., \ion{N}{v}$\lambda$1240, \ion{C}{iv}$\lambda$1549, \ion{He}{ii}$\lambda$1640, and \ion{C}{iii}]$\lambda$1909, are useful tools to measure the metallicity in high-$z$ AGNs, because we can observe these lines in the optical. The metallicity of AGNs has been extensively studied (see Hamann \& Ferland 1999 for a review), in particular by focusing on broad-line region (BLRs). Nagao et al. (2006a) measured emission-line flux ratios of SDSS quasars, e.g., \ion{N}{v}/\ion{C}{iv}, (\ion{Si}{iv}+\ion{O}{iv}])/\ion{C}{iv}, \ion{Al}{iii}/\ion{C}{iv}, and \ion{N}{v}/\ion{He}{ii}, which are sensitive to metallicity. The authors found that there is no redshift evolution of the BLR metallicity in the redshift range of $2.0 < z < 4.5$. Using near-infrared spectroscopic observations of six luminous quasars at $5.8 < z < 6.3$, Jiang et al. (2007) found no strong evolution of the BLR metallicity up to $z \sim 6$. Recently, Juarez et al. (2009) investigated the BLR metallicity of a sample of 30 quasars in the redshift range of $4.0 < z < 6.4$ through the emission-line flux ratio (\ion{Si}{iv}+\ion{O}{iv}])/\ion{C}{iv} and found that the metallicity is very high even in quasars at $z \sim 6$. These results indicate that the major epoch of chemical evolution in AGNs is at $z > 6$. However, AGN broad lines originate in a very small region in galactic nuclei ($R_{\rm BLR} < 1$ pc; e.g., Suganuma et al. 2006), which may have evolved more rapidly and may be not representative of the metallicity in their host galaxies. Moreover, most studies have reported that the BLR metallicity is significantly higher than the solar value ($Z_{\rm BLR} > Z_\odot$; e.g., Hamann \& Ferland 1992; Dietrich et al. 2003; Nagao et al. 2006a; Jiang et al. 2007), reaching as much as $Z_{\rm BLR} \sim 15 Z_\odot$ in the most extreme cases (Baldwin et al. 2003b; Bentz et al. 2004). These extremely high metallicities are very hard to reconcile with galaxy chemical evolutionary models (e.g., Hamann \& Ferland 1993; Ballero et al. 2008) and therefore probably reflect the peculiar evolution of the nuclear region.

To investigate the metallicity on AGN galactic scales, some studies have focussed on the narrow line region (NLR). In contrast to the BLRs, the typical size of the NLRs is comparable to the size of the host galaxies ($R_{\rm NLR} \sim 10^{1-4}$ pc; e.g., Bennert et al. 2006a, 2006b). The mass of the NLRs is about $M_{\rm NLR} \sim 10^{4-8} M_\odot$, which is much higher than that of BLRs ($M_{\rm BLR} \sim 10^{2-4} M_\odot$; see, e.g., Baldwin et al. 2003a). However, the metallicity of the NLR in high-$z$ AGN can be constrained only in type-2 AGNs, whose BLR and strong continuum emission are obscured and therefore allow the detection and careful measurement of the narrow lines. Because there are only a few optically-selected type-2 quasars discovered at $z > 1$, many studies of the NLR metallicity at $z > 1$ have focussed on high-$z$ radio galaxies (HzRGs; e.g., De Breuck et al. 2000; Iwamuro et al. 2003; Nagao et al. 2006b; Humphrey et al. 2008; Matsuoka et al. 2009). By studying the emission-line flux ratios of \ion{N}{v}/\ion{C}{iv} and \ion{N}{v}/\ion{He}{ii}, De Breuck et al. (2000) found that the typical NLR metallicity of HzRGs is roughly $0.4 Z_\odot < Z_{\rm NLR} < 3.0 Z_\odot$. On the contrary, Nagao et al. (2006b) reported that HzRGs do not show any redshift evolution of the NLR metallicity in the redshift range $1.2 < z < 3.8$ when exploiting a metallicity diagnostic diagram involving \ion{C}{iv}, \ion{He}{ii}, and \ion{C}{iii}], which allows one to separate the metallicity and ionization parameter dependencies. Matsuoka et al. (2009) confirmed the absence of any significant metallicity evolution of the NLRs in HzRGs up to $z \sim 4$ by using a sample of 57 HzRGs at $1 < z < 4$. This result suggests that the main epoch of major evolution of the NLR metallicity occurred at even higher redshifts; it is therefore crucial to investigate the metallicity evolution in HzRGs at $z > 4$.

In this paper we focus on the most distant radio galaxy, TN J0924$-$2201 at $z = 5.19$, discovered by Van Breugel et al. (1999). TN J0924$-$2201 is currently the only target that allows us to investigate the NLR metallicity at $z > 5$. In previous studies, Venemans et al. (2004) found that TN J0924$-$2201 is located in a protocluster environment: the density of Ly$\alpha$ emitters around this object is comparable to the density in protoclusters found around HzRGs at $z < 4$ (see also, Overzier et al. 2006). By analyzing the Spitzer data, Seymour et al. (2007) derived the stellar mass of TN J0924$-$2201, i.e., 10$^{11.10} M_\odot$, comparable with the stellar masses of HzRGs at $1 < z < 4$. Klamer et al. (2005) reported that TN J0924$-$2201 may be a young forming galaxy based on its low dust-to-gas ratio (inferred from its $L_{\rm FIR}/L_{\rm CO}$ ratio). This suggests that TN J0924$-$2201 might be undergoing a significant chemical evolution, which can be verified by studying the metallicity of its NLR. However, since previous optical spectroscopic observations of this radio galaxy have identified only the Ly$\alpha$ emission line (i.e., Van Breugel et al. 1999; Venemans et al. 2004), we have obtained a very deep optical spectrum of TN J0924$-$2201 with FOCAS at the Subaru Telescope aimed at detecting additional narrow lines. In the rest of the paper, we adopt a concordance cosmology with ($\Omega_{\rm M}$, $\Omega_\Lambda$) = (0.27, 0.73) and $H_0$ = 71 km s$^{-1}$ Mpc$^{-1}$.

\section{Observation and data reduction} 

We observed TN J0924$-$2201 with FOCAS (the Faint Object Camera And Spectrograph; Kashikawa et al. 2002) at the Subaru Telescope on 2009 Feb. 1. The observation was performed with the 300R dispersion element and the SO58 filter to cover the 5800\AA \ $< \lambda_{\rm obs} <$ 10400\AA , which cover Ly$\alpha$, \ion{N}{v}, \ion{C}{iv}, and \ion{He}{ii} at the redshift of the source. We adopted an on-chip binning mode of $3 \times 1$ with a spatial scale along the slit of $\sim 0\farcs104$ pixel$^{-1}$, before pixel binning, and a spectral dispersion of $\sim 1.34$\AA \ pixel$^{-1}$. With a slit width of 0\farcs8, the resulting spectral resolution was 8.6\AA, measured with the observed sky lines. The typical seeing was $\sim 0\farcs7$. Individual exposure times are 1200s and the total integration time is 18000s.

Standard data reduction procedures were performed by using the available IRAF tasks. A bias was subtracted by using an averaged bias image, and flat-fielding was performed using dome-flat images. Cosmic-ray events were then removed by using the {\tt lacos\_spec} task (van Dokkum 2001). The wavelength calibration was completed by using sky lines, the sky subtraction by nodding the object along the slit and refined using the {\tt background} task. After sky subtraction, we flux-calibrated the spectrum by using observations of the spectrophotometric standard star G191B2B (Oke 1990). After excluding low-$S/N$ frames, we obtained the final stacked two-dimensional spectrum (total exposure time is 12000s) and extracted a single spectrum of the source using an aperture size of 2\farcs18 along the slit. We then corrected for the reddening from the Galactic extinction of $E(B-V) = 0.057$ determined from the dust maps of Schlegel et al. (1998).

\section{Results} 

The final reduced spectrum of TN J0924$-$2201 is shown in Figure 1. In addition to Ly$\alpha$, the \ion{C}{iv} emission is detected with high statistical significance ($S/N \sim 10$). This is the first detection of the atomic emission line from metals in NLR gas clouds at $z > 5$, which allows us to investigate the chemical properties on galactic scales in the early universe. We measured emission-line fluxes and profile widths of Ly$\alpha$ and \ion{C}{iv} by fitting a Gaussian function using the IRAF task {\tt splot} (see Table 1). The line widths after the instrumental broadening correction are 1098 and 426 km s$^{-1}$, respectively. Note that these widths are typical for the NLR emission. For \ion{N}{v} and \ion{He}{ii} lines, 3$\sigma$ upper limits of their fluxes are given by assuming the same line width of \ion{C}{iv}.

\begin{figure} 
\centering
\begin{tabular}{c}
\includegraphics[width=8.5cm]{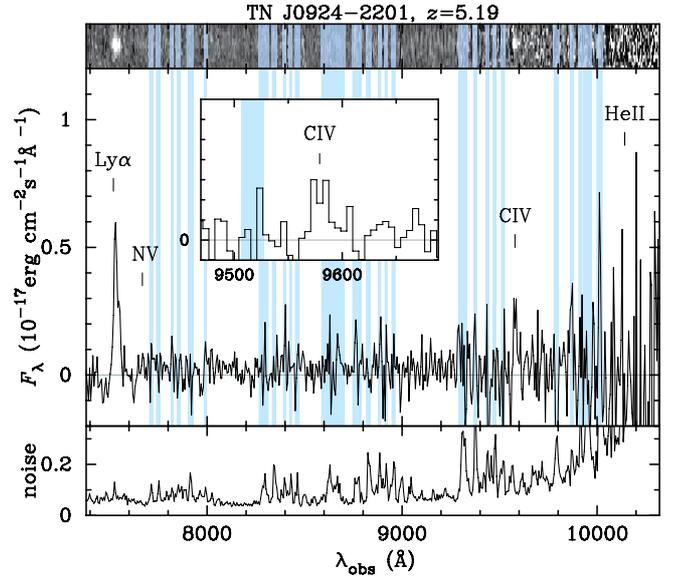}
\end{tabular}
\caption{Final spectrum of TN J0924$-$2201 observed with FOCAS after adopting 4 pixel binning in wavelength direction (the middle panel). The cyan-shaded regions denote the strong sky-line wavelengths. The top panel shows the two-dimensional spectrum. The bottom panel shows the noise spectrum with the same spectral binning.}
\end{figure} 

We calculated the redshifts from the observed wavelengths of Ly$\alpha$ and \ion{C}{iv}. From the Ly$\alpha$ line, the derived redshift is $z = 5.195$, which is intermediate between the two previously measured redshifts of $z = 5.192$ (Van Breugel et al. 1999) and of $z = 5.1989\pm0.0006$ (Venemans et al. 2004). We also derived the redshift from \ion{C}{iv} line, $z = 5.184$, which is lower than that inferred from the Ly$\alpha$ line. The Ly$\alpha$ line profile is likely affected by absorption from the intergalactic medium (IGM), resulting in a artificially higher redshift. Therefore, the redshift derived from \ion{C}{iv} line ($z =5.184$) is more reliable than that from Ly$\alpha$. Note that a possible second feature of the Ly$\alpha$ line is seen on the red wing. This feature may correspond to infalling IGM discussed in Dijkstra et al. (2006).

\begin{table} 
\caption{Emission-line properties of TN J0924$-$2201.}
\begin{tabular}{l c c c c}
\hline \hline \noalign{\smallskip}
\multicolumn{1}{c}{Line ID} & \multicolumn{1}{c}{Flux} & \multicolumn{1}{c}{$\lambda_{\rm c}$$^a$} & \multicolumn{1}{c}{$z$} & \multicolumn{1}{c}{$FWHM^b$}\\
\multicolumn{1}{c}{} & \multicolumn{1}{c}{[10$^{-17}$ erg cm$^{-2}$ s$^{-1}$]} & \multicolumn{1}{c}{[\AA]} & \multicolumn{1}{c}{} & \multicolumn{1}{c}{[\AA]}\\
\noalign{\smallskip} \hline \noalign{\smallskip}
Ly$\alpha$ & 16.10$\pm$0.56 & 7533.2 & 5.195 & 28.9\\
\noalign{\smallskip}
N{\sc v} & $<$ 1.25$^c$ & -- & -- & --\\
\noalign{\smallskip}
C{\sc iv} & 5.46$\pm$0.52 & 9579.2 & 5.184 & 16.1\\
\noalign{\smallskip}
He{\sc ii} & $<$ 5.54$^c$ & -- & -- & --\\
\noalign{\smallskip} \hline
\end{tabular}
$^a$ Central wavelength; $^b$ observed $FWHM$ (includes instrumental broadening); $^c$ 3$\sigma$ upper-limit fluxes.
\end{table} 

\section{Discussion} 

\subsection{Metallicity} 

The detection of the \ion{C}{iv} emission allows us to investigate the NLR metallicity at $z = 5.19$. First, the detection of \ion{C}{iv} indicates that there is plenty of C$^{3+}$ ions even at $z = 5.19$. The CO detection in Klamer et al. (2005) supports the suggestion that TN J0924$-$2201 had already experienced a significant metal enrichment in the NLR gas clouds at $z > 5.19$.

To investigate a redshift evolution of the metallicity, we compared some emission-line flux ratios of TN J0924$-$2201 with those of lower-$z$ HzRGs. We compiled emission-line flux data of HzRGs from the literature (De Breuck et al. 2000; Matsuoka et al. 2009) and plot their Ly$\alpha$/\ion{C}{iv}, \ion{N}{v}/\ion{C}{iv}, and \ion{C}{iv}/\ion{He}{ii} ratios as a function of redshift (Figure 2). Because the emissivity of collisionally-excited emission lines such as \ion{C}{iv} increase at low metallicity (e.g., Nagao et al. 2006b), the Ly$\alpha$/\ion{C}{iv} ratio of TN J0924$-$2201 is expected to be lower than that of lower redshift radio galaxies if there is any significant evolution of the NLR metallicity. As shown in Figure 2, the Ly$\alpha$/\ion{C}{iv} ratio of TN J0924$-$2201 is somewhat lower than the average value observed in radio galaxies at lower redshifts, suggesting some metallicity evolution. However, the Ly$\alpha$/\ion{C}{iv} ratio in TN J0924$-$2201 is still within the scatter observed in lower redshift radio galaxies, which prevents us from making any strong claim on the metallicity evolution based on this line ratio. Moreover, the Ly$\alpha$/\ion{C}{iv} flux ratio is highly uncertain as a consequence of the IGM absorption of Ly$\alpha$, which increases at higher redshift. Another source of uncertainty is that the Ly$\alpha$ flux may be partly associated with star formation in the host galaxy. De Breuck et al. (2000) reported that Ly$\alpha$ is unusually strong in some HzRGs, especially at $z > 3$. Villar-Mart{\'i}n et al. (2007) also confirmed that $z > 3$ radio galaxies tend to show higher Ly$\alpha$ luminosities than lower-$z$ ones. These results suggest that the radio galaxy phenomenon is more often associated with massive starbursts at $z > 3$ than at $z < 3$. However, TN J0924$-$2201 does not follow the trend reported by Villar-Mart{\'i}n et al. (2007). More specifically, the Ly$\alpha$ luminosity of TN J0924$-$2201 is $4.8 \times 10^{43}$ erg s$^{-1}$, comparable with the median of Ly$\alpha$ luminosity of HzRGs at $z < 3$, i.e., $4.4 \times 10^{43}$ erg s$^{-1}$ (Villar-Mart{\'i}n et al. 2007).

The middle panel of Figure 2 shows the emission-line flux ratio of \ion{N}{v}/\ion{C}{iv} as a function of redshift. The \ion{N}{v} line is sometimes regarded as a metallicity indicator for NLRs (e.g., van Ojik et al. 1994; De Breuck et al. 2000; Vernet et al. 2001; Overzier et al. 2001; Humphrey et al. 2008) and therefore it is interesting to examine the redshift dependence of the \ion{N}{v}/\ion{C}{iv} flux ratio as an indicator of the chemical evolution of HzRGs. Our upper limit on the \ion{N}{v}/\ion{C}{iv} ratio is below many detections at low redshift, but also consistent with several other upper limits. Hence our upper limit on \ion{N}{v}/\ion{C}{iv} cannot really indicate whether or not the metallicity of the NLR of TN J0924$-$2201 is lower than observed in lower redshift galaxies.

The bottom panel of Figure 2 shows the emission-line flux ratio of \ion{C}{iv}/\ion{He}{ii} as a function of redshift. Because the luminosity of the \ion{C}{iv} line increases at low metallicity, as mentioned above, this lower limit indicates that the NLR metallicity of TN J0924$-$2201 at $z = 5.19$ is comparable to or lower than that at $z < 5$, as in the case of the \ion{N}{v}/\ion{C}{iv}.

\begin{figure} 
\centering
\begin{tabular}{c}
\includegraphics[width=8.5cm]{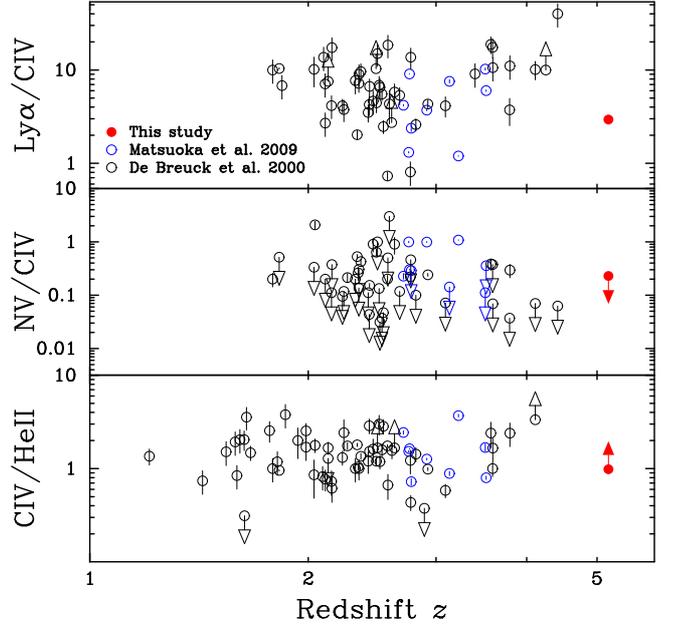}
\end{tabular}
\caption{Flux ratios of Ly$\alpha$/\ion{C}{iv}, \ion{N}{v}/\ion{C}{iv}, and \ion{C}{iv}/\ion{He}{ii} in HzRGs, as a function of redshift. TN J0924$-$2201 is shown with the red-filled circle. The blue-open and black-open circles show the previous data of HzRGs given in Matsuoka et al. (2009) and De Breuck et al. (2000), respectively. The 3$\sigma$ upper and lower limits are shown with arrows.}
\end{figure} 

\subsection{Carbon abundance} 

\begin{figure*}
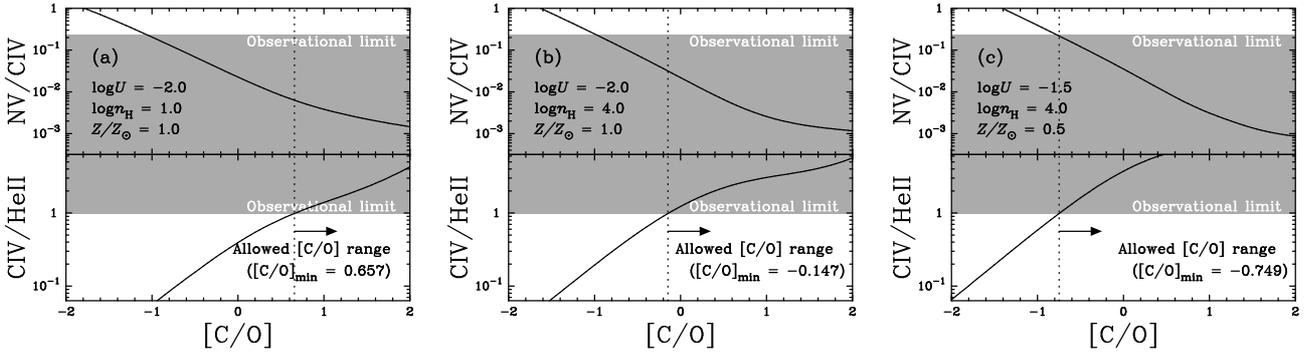
 
\centering
\begin{tabular}{c c c}
\includegraphics[width=5.4cm]{figure3a.ps} & \includegraphics[width=5.4cm]{figure3b.ps} & \includegraphics[width=5.4cm]{figure3c.ps}
\end{tabular}
\caption{Solid lines show the predicted dependencies of the \ion{N}{v}/\ion{C}{iv} and \ion{C}{iv}/\ion{He}{ii} flux ratios on the carbon relative abundance [C/O]. The panels (a), (b), and (c) show models with ($n_{\rm H}$, $U$, $Z_{\rm NLR}/Z_\odot$) = ($10^{1.0}$, $10^{-2.0}$, 1.0), ($10^{4.0}$, $10^{-2.0}$, 1.0), and ($10^{4.0}$, $10^{-1.5}$, 0.5), respectively. The gray shadows denote the ranges that are consistent withe observational limits on the \ion{N}{v}/\ion{C}{iv} and \ion{C}{iv}/\ion{He}{ii} flux ratios in TN J0924$-$2201. The allowed [C/O] range is shown as the dotted line and arrow. The minimum carbon relative abundance ([C/O]$_{\rm min}$) for each model is also given in each panel.}
\end{figure*} 

In this section we use the observational constraints discussed above to investigate the relative carbon abundance in TN J0924$-$2201. The [C/O] abundance ratio is particularly interesting to investigate galaxy evolution in the early universe, since carbon enrichment is delayed compared to $\alpha$ elements because a significant fraction of carbon is produced from intermediate-mass stars, hence [C/O] can be considered a ``clock'' of star formation (e.g., Hamann \& Ferland 1993; Matteucci \& Padovani 1993). In particular, a significant decrease in the carbon relative abundance should be expected at high redshifts if galaxies are indeed in a chemically young phase.

Motivated by this scenario, we examined the dependence of the carbon abundance on the emission-line flux ratios, e.g., \ion{N}{v}/\ion{C}{iv} and \ion{C}{iv}/\ion{He}{ii}, by running photoionization models using the Cloudy (version 08.00; Ferland et al. 1998). We used the ``table AGN'' command for the input SED of ionizing photons. We then assumed NLR clouds, with hydrogen density in the range $n_{\rm H} = 10^{1.0} - 10^{4.0}$ cm$^{-3}$, ionization parameters in the range $U = 10^{-2.5} - 10^{-1.5}$, and (total) metallicity in the range $Z_{\rm NLR} = 0.5 - 2.0 Z_\odot$. These parameter ranges are typical of low-$z$ AGNs (e.g., Nagao et al. 2001, 2002), though it has not been confirmed observationally that these parameter ranges can be adopted for NLRs at $z > 5$. Note that very low or very high metallicities are not expected for the NLR in TN J0924$-$2201, as discussed in Section 4.1. For the chemical composition, we assumed that all metals scale by keeping solar ratios except for C, He, and N. We then adopted the analytical expressions given in Dopita et al. (2000) to obtain the helium and nitrogen relative abundances as a function of metallicity. To investigate the dependence of emission-line flux ratios on the relative carbon abundance, we examined in our model runs the range $-2 <$ [C/O] $< 2$. Dust grains were not included since we consider relatively high-ionization lines (see Nagao et al. 2003).

Figure 3 shows part of our model results. By comparing our observational limits with the model predictions, we infer lower limits of the carbon relative abundance ([C/O]$_{\rm min}$) for each combination of parameters used in the Cloudy runs. We obtain that carbon in TN J0924$-$2201 must have been already significantly enriched. More specifically, in most cases, with realistic NLR physical parameters, the inferred relative carbon abundance is [C/O]$_{\rm min} > -0.5$ (e.g., panels a and b in Figure 3). Only in some extreme cases, with low general metal abundance ($Z = 0.5Z_\odot$) and high ionization parameter ($U = 10^{-1.5}$), the inferred lower limit may be as low [C/O]$_{\rm min} > -0.7$.

The inferred lower limit on the carbon relative abundance ([C/O]$_{\rm min} > -0.5$) is higher than some model predictions expected for the initial phases of galaxy evolution. For instance, Hamann \& Ferland (1993) present the evolution of relative elemental abundance ratios adopting various star-formation histories and IMFs, and all of their models predict [C/O] $< -0.5$ at galaxy ages of $t_{\rm age} \la 100-300$ Myr. Adopting their models, our observational limit on [C/O] then suggests that the age of TN J0924$-$2201 must be at least of a few hundred Myr; i.e., according to our analysis TN J0924$-$2201 is not in the expected extremely young phase.

\section{Conclusion} 

To investigate the chemical evolution of HzRGs at $z > 5$, we investigated the rest-frame UV spectrum of TN J0924$-$2201 and obtained the following main results.
\begin{enumerate}
\item
We detected \ion{C}{iv} emission in TN J0924$-$2201 at $z = 5.19$; this is the first detection of metal emission lines from NLR gas clouds at $z > 5$, suggesting that a significant amount of carbon exists even at $z > 5$ in the early universe.
\item
The measured Ly$\alpha$/\ion{C}{iv} flux ratio is lower than that of HzRGs at $3 < z < 5$; this can possibly be ascribed to weaker star-formation activity and/or heavier IGM absorption on Ly$\alpha$.
\item
By comparing observational limits on emission-line flux ratios (and in particular \ion{N}{v}/\ion{C}{iv} and \ion{C}{iv}/\ion{He}{ii}) with photoionization models, we inferred that the carbon is already relatively high in TN J0924$-$2201, more specifically [C/O] $> -0.5$, suggesting that TN J0924$-$2201 is not in an extremely young phase, but likely already a few hundred million yrs old.
\end{enumerate}

\begin{acknowledgements} 

We thank Yuko Ideue and the Subaru Telescope staff for the assisting our FOCAS observation. K.M. acknowledges financial support from the Japan Society for the Promotion of Science (JSPS). Data analysis were in part carried out on common-use data analysis computer system at the Astronomy Data Center, ADC, of the National Astronomical Observatory of Japan. T.N. and Y.T. are financially supported by JSPS (Grant Nos. 23244031 and 23654068).

\end{acknowledgements} 


\end{document}